\documentstyle[prl,twocolumn,aps]{revtex}

\def\temp{1.35}%
\let\tempp=\relax
\expandafter\ifx\csname psboxversion\endcsname\relax
\else
    \ifdim\temp cm>\psboxversion cm
    \else
      \let\temp=\psboxversion
      \let\tempp= 
    \fi
\fi
\tempp
\let\psboxversion=\temp
\catcode`\@=11
\def\psfortextures{
\def\PSspeci@l##1##2{%
\special{illustration ##1\space scaled ##2}%
}}%
\def\psfordvitops{
\def\PSspeci@l##1##2{%
\special{dvitops: import ##1\space \the\drawingwd \the\drawinght}%
}}%
\def\psfordvips{
\def\PSspeci@l##1##2{%
\d@my=0.1bp \d@mx=\drawingwd \divide\d@mx by\d@my
\includegraphics{##1\space}}}%
\def\psforoztex{
\def\PSspeci@l##1##2{%
\special{##1 \space
      ##2 1000 div dup scale
      \number-\psllx\space\space \number-\pslly\space\space translate
}}}%
\def\psfordvitps{
\def\dvitpsLiter@ldim##1{\dimen0=##1\relax
\special{dvitps: Literal "\number\dimen0\space"}}%
\def\PSspeci@l##1##2{%
\at(0bp;\drawinght){%
\special{dvitps: Include0 "psfig.psr"}
\dvitpsLiter@ldim{\drawingwd}%
\dvitpsLiter@ldim{\drawinght}%
\dvitpsLiter@ldim{\psllx bp}%
\dvitpsLiter@ldim{\pslly bp}%
\dvitpsLiter@ldim{\psurx bp}%
\dvitpsLiter@ldim{\psury bp}%
\special{dvitps: Literal "startTexFig"}%
\special{dvitps: Include1 "##1"}%
\special{dvitps: Literal "endTexFig"}%
}}}%
\def\psfordvialw{
\def\PSspeci@l##1##2{
\special{language "PostScript",
position = "bottom left",
literal "  \psllx\space \pslly\space translate
  ##2 1000 div dup scale
  -\psllx\space -\pslly\space translate",
include "##1"}
}}%
\def\psforptips{
\def\PSspeci@l##1##2{{
\d@mx=\psurx bp
\advance \d@mx by -\psllx bp
\divide \d@mx by 1000\multiply\d@mx by \xscale
\incm{\d@mx}
\let\tmpx\dimincm
\d@my=\psury bp
\advance \d@my by -\pslly bp
\divide \d@my by 1000\multiply\d@my by \xscale
\incm{\d@my}
\let\tmpy\dimincm
\d@mx=-\psllx bp
\divide \d@mx by 1000\multiply\d@mx by \xscale
\d@my=-\pslly bp
\divide \d@my by 1000\multiply\d@my by \xscale
\at(\d@mx;\d@my){\special{ps:##1 x=\tmpx cm, y=\tmpy cm}}
}}}%
\def\psonlyboxes{
\def\PSspeci@l##1##2{%
\at(0cm;0cm){\boxit{\vbox to\drawinght
  {\vss\hbox to\drawingwd{\at(0cm;0cm){\hbox{({\tt##1})}}\hss}}}}
}}%
\def\psloc@lerr#1{%
\let\savedPSspeci@l=\PSspeci@l%
\def\PSspeci@l##1##2{%
\at(0cm;0cm){\boxit{\vbox to\drawinght
  {\vss\hbox to\drawingwd{\at(0cm;0cm){\hbox{({\tt##1}) #1}}\hss}}}}
\let\PSspeci@l=\savedPSspeci@l
}}%
\newread\pst@mpin
\newdimen\drawinght\newdimen\drawingwd
\newdimen\psxoffset\newdimen\psyoffset
\newbox\drawingBox
\newcount\xscale \newcount\yscale \newdimen\pscm\pscm=1cm
\newdimen\d@mx \newdimen\d@my
\newdimen\pswdincr \newdimen\pshtincr
\let\ps@nnotation=\relax
{\catcode`\|=0 |catcode`|\=12 |catcode`|
|catcode`#=12 |catcode`*=14
|xdef|backslashother{\}*
|xdef|percentother{
|xdef|tildeother{~}*
|xdef|sharpother{#}*
}%
\def\R@moveMeaningHeader#1:->{}%
\def\uncatcode#1{%
\edef#1{\expandafter\R@moveMeaningHeader\meaning#1}}%
\def\execute#1{#1}
\def\psm@keother#1{\catcode`#112\relax}
\def\executeinspecs#1{%
\execute{\begingroup\let\do\psm@keother\dospecials\catcode`\^^M=9#1\endgroup}}%
\def\@mpty{}%
\def\matchexpin#1#2{
  \fi%
  \edef\tmpb{{#2}}%
  \expandafter\makem@tchtmp\tmpb%
  \edef\tmpa{#1}\edef\tmpb{#2}%
  \expandafter\expandafter\expandafter\m@tchtmp\expandafter\tmpa\tmpb\endm@tch%
  \if\match%
}%
\def\matchin#1#2{%
  \fi%
  \makem@tchtmp{#2}%
  \m@tchtmp#1#2\endm@tch%
  \if\match%
}%
\def\makem@tchtmp#1{\def\m@tchtmp##1#1##2\endm@tch{%
  \def\tmpa{##1}\def\tmpb{##2}\let\m@tchtmp=\relax%
  \ifx\tmpb\@mpty\def\match{YN}%
  \else\def\match{YY}\fi%
}}%
\def\incm#1{{\psxoffset=1cm\d@my=#1
 \d@mx=\d@my
  \divide\d@mx by \psxoffset
  \xdef\dimincm{\number\d@mx.}
  \advance\d@my by -\number\d@mx cm
  \multiply\d@my by 100
 \d@mx=\d@my
  \divide\d@mx by \psxoffset
  \edef\dimincm{\dimincm\number\d@mx}
  \advance\d@my by -\number\d@mx cm
  \multiply\d@my by 100
 \d@mx=\d@my
  \divide\d@mx by \psxoffset
  \xdef\dimincm{\dimincm\number\d@mx}
}}%
\newif\ifNotB@undingBox
\newhelp\PShelp{Proceed: you'll have a 5cm square blank box instead of
your graphics.}%
\def\s@tsize#1 #2 #3 #4\@ndsize{
  \def\psllx{#1}\def\pslly{#2}%
  \def\psurx{#3}\def\psury{#4}
  \ifx\psurx\@mpty\NotB@undingBoxtrue
  \else
    \drawinght=#4bp\advance\drawinght by-#2bp
    \drawingwd=#3bp\advance\drawingwd by-#1bp
  \fi
  }%
\def\sc@nBBline#1:#2\@ndBBline{\edef\p@rameter{#1}\edef\v@lue{#2}}%
\def\g@bblefirstblank#1#2:{\ifx#1 \else#1\fi#2}%
{\catcode`\%=12
\xdef\B@undingBox{
\def\ReadPSize#1{
 \readfilename#1\relax
 \let\PSfilename=\lastreadfilename
 \openin\pst@mpin=#1\relax
 \ifeof\pst@mpin \errhelp=\PShelp
   \errmessage{I haven't found your postscript file (\PSfilename)}%
   \psloc@lerr{was not found}%
   \s@tsize 0 0 142 142\@ndsize
   \closein\pst@mpin
 \else
   \if\matchexpin{\GlobalInputList}{, \lastreadfilename}%
   \else\xdef\GlobalInputList{\GlobalInputList, \lastreadfilename}%
     \immediate\write\psbj@inaux{\lastreadfilename,}%
   \fi%
   \loop
     \executeinspecs{\catcode`\ =10\global\read\pst@mpin to\n@xtline}%
     \ifeof\pst@mpin
       \errhelp=\PShelp
       \errmessage{(\PSfilename) is not an Encapsulated PostScript File:
           I could not find any \B@undingBox: line.}%
       \edef\v@lue{0 0 142 142:}%
       \psloc@lerr{is not an EPSFile}%
       \NotB@undingBoxfalse
     \else
       \expandafter\sc@nBBline\n@xtline:\@ndBBline
       \ifx\p@rameter\B@undingBox\NotB@undingBoxfalse
         \edef\t@mp{%
           \expandafter\g@bblefirstblank\v@lue\space\space\space}%
         \expandafter\s@tsize\t@mp\@ndsize
       \else\NotB@undingBoxtrue
       \fi
     \fi
   \ifNotB@undingBox\repeat
   \closein\pst@mpin
 \fi
\message{#1}%
}%
\def\psboxto(#1;#2)#3{\vbox{%
   \ReadPSize{#3}%
   \advance\pswdincr by \drawingwd
   \advance\pshtincr by \drawinght
   \divide\pswdincr by 1000
   \divide\pshtincr by 1000
   \d@mx=#1
   \ifdim\d@mx=0pt\xscale=1000
         \else \xscale=\d@mx \divide \xscale by \pswdincr\fi
   \d@my=#2
   \ifdim\d@my=0pt\yscale=1000
         \else \yscale=\d@my \divide \yscale by \pshtincr\fi
   \ifnum\yscale=1000
         \else\ifnum\xscale=1000\xscale=\yscale
                    \else\ifnum\yscale<\xscale\xscale=\yscale\fi
              \fi
   \fi
   \divide\drawingwd by1000 \multiply\drawingwd by\xscale
   \divide\drawinght by1000 \multiply\drawinght by\xscale
   \divide\psxoffset by1000 \multiply\psxoffset by\xscale
   \divide\psyoffset by1000 \multiply\psyoffset by\xscale
   \global\divide\pscm by 1000
   \global\multiply\pscm by\xscale
   \multiply\pswdincr by\xscale \multiply\pshtincr by\xscale
   \ifdim\d@mx=0pt\d@mx=\pswdincr\fi
   \ifdim\d@my=0pt\d@my=\pshtincr\fi
   \message{scaled \the\xscale}%
 \hbox to\d@mx{\hss\vbox to\d@my{\vss
   \global\setbox\drawingBox=\hbox to 0pt{\kern\psxoffset\vbox to 0pt{%
      \kern-\psyoffset
      \PSspeci@l{\PSfilename}{\the\xscale}%
      \vss}\hss\ps@nnotation}%
   \global\wd\drawingBox=\the\pswdincr
   \global\ht\drawingBox=\the\pshtincr
   \global\drawingwd=\pswdincr
   \global\drawinght=\pshtincr
   \baselineskip=0pt
   \copy\drawingBox
 \vss}\hss}%
  \global\psxoffset=0pt
  \global\psyoffset=0pt
  \global\pswdincr=0pt
  \global\pshtincr=0pt 
  \global\pscm=1cm 
}}%
\def\psboxscaled#1#2{\vbox{%
  \ReadPSize{#2}%
  \xscale=#1
  \message{scaled \the\xscale}%
  \divide\pswdincr by 1000 \multiply\pswdincr by \xscale
  \divide\pshtincr by 1000 \multiply\pshtincr by \xscale
  \divide\psxoffset by1000 \multiply\psxoffset by\xscale
  \divide\psyoffset by1000 \multiply\psyoffset by\xscale
  \divide\drawingwd by1000 \multiply\drawingwd by\xscale
  \divide\drawinght by1000 \multiply\drawinght by\xscale
  \global\divide\pscm by 1000
  \global\multiply\pscm by\xscale
  \global\setbox\drawingBox=\hbox to 0pt{\kern\psxoffset\vbox to 0pt{%
     \kern-\psyoffset
     \PSspeci@l{\PSfilename}{\the\xscale}%
     \vss}\hss\ps@nnotation}%
  \advance\pswdincr by \drawingwd
  \advance\pshtincr by \drawinght
  \global\wd\drawingBox=\the\pswdincr
  \global\ht\drawingBox=\the\pshtincr
  \global\drawingwd=\pswdincr
  \global\drawinght=\pshtincr
  \baselineskip=0pt
  \copy\drawingBox
  \global\psxoffset=0pt
  \global\psyoffset=0pt
  \global\pswdincr=0pt
  \global\pshtincr=0pt 
  \global\pscm=1cm
}}%
\def\psbox#1{\psboxscaled{1000}{#1}}%
\newif\ifn@teof\n@teoftrue
\newif\ifc@ntrolline
\newif\ifmatch
\newread\j@insplitin
\newwrite\j@insplitout
\newwrite\psbj@inaux
\immediate\openout\psbj@inaux=psbjoin.aux
\immediate\write\psbj@inaux{\string\joinfiles}%
\immediate\write\psbj@inaux{\jobname,}%
\def\toother#1{\ifcat\relax#1\else\expandafter%
  \toother@ux\meaning#1\endtoother@ux\fi}%
\def\toother@ux#1 #2#3\endtoother@ux{\def\tmp{#3}%
  \ifx\tmp\@mpty\def\tmp{#2}\let\next=\relax%
  \else\def\next{\toother@ux#2#3\endtoother@ux}\fi%
\next}%
\let\readfilenamehook=\relax
\def\re@d{\expandafter\re@daux}
\def\re@daux{\futurelet\nextchar\stopre@dtest}%
\def\re@dnext{\xdef\lastreadfilename{\lastreadfilename\nextchar}%
  \afterassignment\re@d\let\nextchar}%
\def\stopre@d{\egroup\readfilenamehook}%
\def\stopre@dtest{%
  \ifcat\nextchar\relax\let\nextread\stopre@d
  \else
    \ifcat\nextchar\space\def\nextread{%
      \afterassignment\stopre@d\chardef\nextchar=`}%
    \else\let\nextread=\re@dnext
      \toother\nextchar
      \edef\nextchar{\tmp}%
    \fi
  \fi\nextread}%
\def\readfilename{\bgroup%
  \let\\=\backslashother \let\%=\percentother \let\~=\tildeother
  \let\#=\sharpother \xdef\lastreadfilename{}%
  \re@d}%
\xdef\GlobalInputList{\jobname}%
\def\psnewinput{%
  \def\readfilenamehook{
    \if\matchexpin{\GlobalInputList}{, \lastreadfilename}%
    \else\xdef\GlobalInputList{\GlobalInputList, \lastreadfilename}%
      \immediate\write\psbj@inaux{\lastreadfilename,}%
    \fi%
    \let\readfilenamehook=\relax%
    \ps@ldinput\lastreadfilename\relax%
  }\readfilename%
}%
\expandafter\ifx\csname @@input\endcsname\relax    
  \immediate\let\ps@ldinput=\input\def\input{\psnewinput}%
\else
  \immediate\let\ps@ldinput=\@@input
  \def\@@input{\psnewinput}%
\fi%
\def\nowarnopenout{%
 \def\warnopenout##1##2{%
   \readfilename##2\relax
   \message{\lastreadfilename}%
   \immediate\openout##1=\lastreadfilename\relax}}%
\def\warnopenout#1#2{%
 \readfilename#2\relax
 \def\t@mp{TrashMe,psbjoin.aux,psbjoint.tex,}\uncatcode\t@mp
 \if\matchexpin{\t@mp}{\lastreadfilename,}%
 \else
   \immediate\openin\pst@mpin=\lastreadfilename\relax
   \ifeof\pst@mpin
     \else
     \edef\tmp{{If the content of this file is precious to you, this
is your last chance to abort (ie press x or e) and rename it before
retexing (\jobname). If you're sure there's no file
(\lastreadfilename) in the directory of (\jobname), then go on: I'm
simply worried because you have another (\lastreadfilename) in some
directory I'm looking in for inputs...}}%
     \errhelp=\tmp
     \errmessage{I may be about to replace your file named \lastreadfilename}%
   \fi
   \immediate\closein\pst@mpin
 \fi
 \message{\lastreadfilename}%
 \immediate\openout#1=\lastreadfilename\relax}%
{\catcode`\%=12\catcode`\*=14
\gdef\splitfile#1{*
 \readfilename#1\relax
 \immediate\openin\j@insplitin=\lastreadfilename\relax
 \ifeof\j@insplitin
   \message{! I couldn't find and split \lastreadfilename!}*
 \else
   \immediate\openout\j@insplitout=TrashMe
   \message{< Splitting \lastreadfilename\space into}*
   \loop
     \ifeof\j@insplitin
       \immediate\closein\j@insplitin\n@teoffalse
     \else
       \n@teoftrue
       \executeinspecs{\global\read\j@insplitin to\spl@tinline\expandafter
         \ch@ckbeginnewfile\spl@tinline
       \ifc@ntrolline
       \else
         \toks0=\expandafter{\spl@tinline}*
         \immediate\write\j@insplitout{\the\toks0}*
       \fi
     \fi
   \ifn@teof\repeat
   \immediate\closeout\j@insplitout
 \fi\message{>}*
}*
\gdef\ch@ckbeginnewfile#1
 \def\t@mp{#1}*
 \ifx\@mpty\t@mp
   \def\t@mp{#3}*
   \ifx\@mpty\t@mp
     \global\c@ntrollinefalse
   \else
     \immediate\closeout\j@insplitout
     \warnopenout\j@insplitout{#2}*
     \global\c@ntrollinetrue
   \fi
 \else
   \global\c@ntrollinefalse
 \fi}*
\gdef\joinfiles#1\into#2{*
 \message{< Joining following files into}*
 \warnopenout\j@insplitout{#2}*
 \message{:}*
 {*
 \edef\w@##1{\immediate\write\j@insplitout{##1}}*
\w@{
\w@{
\w@{
\w@{
\w@{
\w@{
\w@{
\w@{
\w@{
\w@{
\w@{\string\input\space psbox.tex}*
\w@{\string\splitfile{\string\jobname}}*
\w@{\string\let\string\autojoin=\string\relax}*
}*
 \expandafter\tre@tfilelist#1, \endtre@t
 \immediate\closeout\j@insplitout
 \message{>}*
}*
\gdef\tre@tfilelist#1, #2\endtre@t{*
 \readfilename#1\relax
 \ifx\@mpty\lastreadfilename
 \else
   \immediate\openin\j@insplitin=\lastreadfilename\relax
   \ifeof\j@insplitin
     \errmessage{I couldn't find file \lastreadfilename}*
   \else
     \message{\lastreadfilename}*
     \immediate\write\j@insplitout{
     \executeinspecs{\global\read\j@insplitin to\oldj@ininline}*
     \loop
       \ifeof\j@insplitin\immediate\closein\j@insplitin\n@teoffalse
       \else\n@teoftrue
         \executeinspecs{\global\read\j@insplitin to\j@ininline}*
         \toks0=\expandafter{\oldj@ininline}*
         \let\oldj@ininline=\j@ininline
         \immediate\write\j@insplitout{\the\toks0}*
       \fi
     \ifn@teof
     \repeat
   \immediate\closein\j@insplitin
   \fi
   \tre@tfilelist#2, \endtre@t
 \fi}*
}%
\def\autojoin{%
 \immediate\write\psbj@inaux{\string\into{psbjoint.tex}}%
 \immediate\closeout\psbj@inaux
 \expandafter\joinfiles\GlobalInputList\into{psbjoint.tex}%
}%
\def\centinsert#1{\midinsert\line{\hss#1\hss}\endinsert}%
\def\psannotate#1#2{\vbox{%
  \def\ps@nnotation{#2\global\let\ps@nnotation=\relax}#1}}%
\def\pscaption#1#2{\vbox{%
   \setbox\drawingBox=#1
   \copy\drawingBox
   \vskip\baselineskip
   \vbox{\hsize=\wd\drawingBox\setbox0=\hbox{#2}%
     \ifdim\wd0>\hsize
       \noindent\unhbox0\tolerance=5000
    \else\centerline{\box0}%
    \fi
}}}%
\def\at(#1;#2)#3{\setbox0=\hbox{#3}\ht0=0pt\dp0=0pt
  \rlap{\kern#1\vbox to0pt{\kern-#2\box0\vss}}}%
\newdimen\gridht \newdimen\gridwd
\def\gridfill(#1;#2){%
  \setbox0=\hbox to 1\pscm
  {\vrule height1\pscm width.4pt\leaders\hrule\hfill}%
  \gridht=#1
  \divide\gridht by \ht0
  \multiply\gridht by \ht0
  \gridwd=#2
  \divide\gridwd by \wd0
  \multiply\gridwd by \wd0
  \advance \gridwd by \wd0
  \vbox to \gridht{\leaders\hbox to\gridwd{\leaders\box0\hfill}\vfill}}%
\def\fillinggrid{\at(0cm;0cm){\vbox{%
  \gridfill(\drawinght;\drawingwd)}}}%
\def\textleftof#1:{%
  \setbox1=#1
  \setbox0=\vbox\bgroup
    \advance\hsize by -\wd1 \advance\hsize by -2em}%
\def\textrightof#1:{%
  \setbox0=#1
  \setbox1=\vbox\bgroup
    \advance\hsize by -\wd0 \advance\hsize by -2em}%
\def\endtext{%
  \egroup
  \hbox to \hsize{\valign{\vfil##\vfil\cr%
\box0\cr%
\noalign{\hss}\box1\cr}}}%
\def\frameit#1#2#3{\hbox{\vrule width#1\vbox{%
  \hrule height#1\vskip#2\hbox{\hskip#2\vbox{#3}\hskip#2}%
        \vskip#2\hrule height#1}\vrule width#1}}%
\def\boxit#1{\frameit{0.4pt}{0pt}{#1}}%
\catcode`\@=12 
\psfordvips   

\begin{document}

\draft

\title{Consequence of superfluidity on the expansion of a 
rotating Bose-Einstein condensate}
\author{Mark Edwards$^{a,b}$, Charles W.\ Clark$^b$, P.\ Pedri$^c$, 
L.\ Pitaevskii$^{c,d}$ and S.\ Stringari$^{c}$}
\address{$^{a}$Department of Physics, Georgia Southern University,
Statesboro, GA 30460--8031 USA}
\address{$^{b}$National Institute of Standards and Technology, 
Gaithersburg, MD 20899-8410, USA,}
\address{$^{c}$Dipartimento di Fisica, Universit\`{a} di Trento,
and Istituto Nazionale per la Fisica della Materia, I--38050 Povo, Italy}
\address{$^{d}$ Kapitza Institute for Physical Problems, 
ul.\ Kosygina 2, 117334 Moscow, Russia}

\maketitle

\begin{abstract}
We study the time evolution of a rotating condensate, that 
expands after being suddenly released from the confining trap, by 
solving the hydrodynamic equations of irrotational superfluids.  
For slow initial rotation speeds, $\Omega_{0}$, we find that the 
condensate's angular velocity increases rapidly to a maximum value
and this is accompanied by a minimum in the deformation of the 
condensate in the rotating plane.  During the expansion the sample 
makes a global rotation of approximately $\pi/2$, where the exact value 
depends on $\Omega_{0}$.  This minimum deformation can serve as an
easily detectable signature of superfluidity in a Bose--Einstein 
condensate.
\end{abstract}

\pacs{PACS Numbers: 3.75.Fi, 67.40.Db, 67.90.+Z}


\narrowtext

Recent experiments on trapped atomic Bose gases at very low temperature
have revealed the existence of remarkable superfluid effects.  Initial
evidence of superfluidity came from the observation of quantized
vortices \cite{jila,ens} and of the reduction of dissipative phenomena
\cite{mit}. 

Important signatures of superfluidity are also provided by the study 
of rotating condensates at low angular velocities, in the absence of
quantized vortices, where the irrotationality of the flow implies the
condition $\nabla\times{\bf v} = 0$ everywhere. In particular the quenching
of the moment of inertia provides direct evidence of superfluidity, 
being related to the suppression of the transverse response of  the
many-body system~\cite{baym+}. The recent observation of the scissors
mode~\cite{foot} has provided a clear example of this effect,
confirming with high accuracy the predictions of theory~\cite{guery}.
The purpose of the present work is to investigate the consequences of
superfluidity on the expansion of an initially rotating and vortex--free
condensate. 

In the absence of rotation the expansion of a condensate after releasing
the confining trap is well understood~\cite{castin,LVM}. In particular the
shape of a condensate, initially asymmetric in the $x-y$ plane, when 
released approaches an asymptotic configuration where the role of the long 
and short axes is inverted with respect to the initial geometry. This 
inversion occurs because of the larger pressure felt by the system in the 
short direction where the gradient of the density is higher.  As a 
consequence the cross section of the condensate in the $x-y$ plane will be 
instantaneously circular at some intermediate time during the expansion. 
The time needed to reach this symmetrical configuration depends both on 
the initial deformation and on the presence of interatomic forces. For a 
noninteracting gas this time is given by $(\omega_{x}\omega _{y})^{-1/2}$ 
where $\omega_{x}$, $\omega_y$ are the trapping frequencies of the harmonic 
potential which initially confines the gas.  Interactions modify the time 
scale of the expansion in a profound way. In the Thomas--Fermi regime 
and in the limit of a very elongated condensate ($\omega_{y}\ll\omega_{x}$) 
the time needed to reach symmetry is given by $1/\omega_y$. In this case 
the long axis does not expand significantly and the symmetry of the 
configuration is achieved because of the fast expansion of the short 
axis~\cite{castin}.

If the condensate is initially rotating new features emerge during the
expansion which are the object of the present investigation. In particular
we predict that the condensate will never reach the symmetric configuration,
because of the occurrence of a repulsive barrier caused by the
irrotationality of the flow.

An estimate of the effect is easily obtained by using angular momentum and
energy conservation. Angular momentum conservation implies that when the
system approaches the symmetric configuration in the $x-y$ plane its angular
velocity in the same plane becomes larger and larger as a consequence of the
vanishing of the moment of inertia. This is a crucial consequence of the
irrotationality of the superfluid motion. On the other hand the angular
velocity cannot increase too much because of energy conservation. By
assuming that the initial energy of the system $E_{0}$ is mostly converted
into rotational energy, one finds $E_{0}\sim\Omega_{cr}^{2}\Theta_{cr}/2$. 
On the other hand angular momentum conservation requires 
$\Omega_{0}\Theta _{0}=\Omega _{cr}\Theta _{cr}$ where $\Omega _{0} $ and 
$\Theta_{0}$ are the angular velocity and the moment of inertia at $t=0$, 
while $\Omega _{cr}$ and $\Theta _{cr}$ are the corresponding values at the 
point of maximum rotational energy (hereafter called critical time). One 
finds: $\Theta _{cr}\sim \Omega _{0}^{2}\Theta _{0}^{2}/E_{0}$ and 
$\Omega _{cr}\sim E_{0}/(\Omega _{0}\Theta _{0})$. Since in a superfluid 
the moment of inertia is given by the irrotational value 
$\Theta = \delta^{2}\Theta _{rig}$, where 
\begin{equation}
\delta ={\frac{<y^{2}-x^{2}>}{<y^{2}+x^{2}>}}
\label{delta}
\end{equation}
is the deformation parameter of the cloud calculated with respect to its
symmetry axes, and $\Theta _{rig}=Nm<x^{2}+y^{2}>$
is the rigid value of the moment of inertia, one concludes that the system
will reach a minimum deformation which depends linearly on the initial
angular velocity $\Omega _{0}$, while its angular velocity, at the same
time, will behave like $1/\Omega _{0}$.

The previous discussion suggests the following scenario: the rotating 
system will first expand in the short direction (we consider here a 
condensate significantly deformed in the $x-y$ plane at $t=0$).  At a 
time, when the deformation nears its minimum value, the condensate will 
rotate rapidly causing the inversion of the long and short axes.  The
condensate then continues to expand along the original long axis.  The 
occurrence of a minimum deformation during the exansion and the 
corresponding increase of the angular velocity can be regarded as a 
signature of the superfluid behavior of the system. In fact a classical 
rotating gas expands in a very different way, because of the absence of 
the irrotationality constraint. In the classical case, neglecting the 
effects of collisions during the expansion, the angular velocity of the 
sample will decrease smoothly as a function of time according to the law  
$\Omega_{cl}(t) = \Omega_{0}/(1 + \Omega_{0}^{2}t^{2})$.

In the following we will provide a quantitative description of the 
expansion of a rotating Bose-Einstein condensate at zero temperature
in the Thomas--Fermi regime.  This regime is relevant to most current 
experimental situations.  The opposite limit, corresponding to the 
exapansion of a rotating ideal gas,  will be considered in a separate
paper. In the Thomas--Fermi regime the equations of motion take the
simplified form of the hydrodynamic equations of superfluids~\cite{RMP}. 
For the free expansion these equations, valid in the laboratory frame, 
can be written as an equation of continuity 
\begin{equation}
\frac{\partial n}{\partial t} +
{\bf\nabla}\cdot\left(n{\bf v}\right) = 0,
\label{continuity}
\end{equation}
and a Newton's 2nd--law ``force'' equation
\begin{equation}
m\frac{\partial {\bf v}}{\partial t}+{\bf \nabla }\left( \frac{1}{2}%
mv^{2} + gn\right) = 0.
\label{force}
\end{equation}
where $n$ and ${\bf v}$ are the condensate density and velocity field,
respectively, and $m$ is the atomic mass. The factor 
$g=4\pi\hbar^{2}a_{sc}/m$ in the second equation measures the strength 
of the atom--atom interaction where $a_{sc}$ is the $s$--wave scattering 
length.

The condensate is assumed to be held at equilibrium in a rotating trap just
before turning off the external potential. In~\cite{RZS} it was shown that
in the rotating frame the stationary solution of the equations of motion
still corresponds to a spheroidal form for the density profile. At $t=0$ 
we can write 
\begin{equation}
n\left({\bf r},0\right) = 
b_{0} - a_{x0}x^{2} - a_{y0}y^{2} - a_{z0}z^{2}
\label{rho0}
\end{equation}
and 
\begin{equation}
{\bf v}\left({\bf r},0\right) =
\alpha_{0}\nabla\left( xy \right) \ ,
\label{v0}
\end{equation}
where the $x-y$ plane is perpendicular to the axis of rotation.
The coefficients $b_{0}$, $a_{i0}$ and $\alpha_{0}$ are determined by the
number of atoms, $N$, the trap frequencies, $\omega_{i}$, and the
initial angular velocity, $\Omega_{0}$, and should fulfil the stationary 
criteria discussed in~\cite{RZS}. For example one has the relationship 
\begin{equation}
\alpha_{0}= 
\left(
\frac{a_{x0} - a_{y0}}{a_{x0} + a_{y0}} 
\right)\Omega_{0}
\, .
\label{alpha0} 
\end{equation}

If the angular velocity $\Omega _{0}$ is small, $b_{0}$ and $a_{i0}$
coincide with their values in the absence of rotation andare given by 
$b_{0}=\mu_{TF}/g$ and $a_{i0}=m\omega_{i}^{2}/2g$ , where $\mu _{TF}$ 
is the Thomas--Fermi value of the chemical potential. In the same limit 
of small angular velocity the chemical potential fixes the initial radii, 
$X_{i0}=(b_{0}/a_{i0})^{1/2}$ (see Eq.\ (\ref{rho0})) of the condensate
according to the relationships $2\mu _{TF}=m\omega_{x}^{2}X_{0}^{2}= m
\omega_{y}^{2}Y_{0}^{2}=m\omega_{z}^{2}Z_{0}^{2}$.

The solution of the expanding condensate can be written in the form 
\begin{eqnarray}
n\left({\bf r},t\right) &=&
b\left(t\right) - 
a_{x}\left(t\right)x^{2} - 
a_{y}\left(t\right)y^{2} -
a_{z}\left(t\right)z^{2}\nonumber\\
&-&
a\left(t\right)xy  
\label{rho_ansatz}
\end{eqnarray}
and 
\begin{eqnarray}
{\bf v}\left( {\bf r},t\right) &=&
\frac{1}{2}{\bf \nabla }
\Big(
\alpha_{x}\left(t\right)x^{2} + 
\alpha_{y}\left(t\right)y^{2} +
\alpha_{z}\left(t\right)z^{2}\nonumber\\
&+&
2\alpha\left(t\right)xy
\Big) \, .  
\label{v_ansatz}
\end{eqnarray}
The condensate occupies the region in space where $n>0$. Note that the
function $b(t)$ defines the density at the center of the trap, 
$n\left(0,t\right) = b\left(t\right) $. At $t=0$ one has 
$b\left(0\right)=b_{0}, a_{i}\left(0\right) = a_{i0},\alpha\left(0\right) 
=\alpha_{0}$ and $a\left(0\right) = \alpha_{i}\left(0\right) = 0$. 

Equation (\ref{rho_ansatz}) shows that during the expansion  the density
 keeps a
spheroidal form. Because of the initial
angular velocity in the $x-y$ plane the symmetry axis will rotate and the
density will be no longer diagonal in the $x$, $y$ coordinates. The angle of
rotation can be easily evaluated in terms of the  coefficients of 
(\ref{rho_ansatz}) by introducing the usual
transformation 
$x=x^{\prime}\cos \phi +y^{\prime}\sin \phi$,
$y=-x^{\prime}\sin \phi +y^{\prime}\cos \phi$ and $z = z^{\prime}$.
This transformation diagonalizes 
the quantity $s^{2}= a_{x}\left( t\right)
x^{2}+a_{y}\left( t\right) y^{2}+a_{z}\left( t\right) z^{2}+a\left( t\right)
xy$ which takes the form  
\begin{eqnarray}
s^{2}&=&{\frac{1}{2}}(a_{x}+a_{y})((x^{\prime})^{2}+(y^{\prime})^{2})
+{\frac{1}{2}}
((a_{x}-a_{y})\cos 2\phi\nonumber \\
&-&a\sin 2\phi )((x^{\prime})^{2}-(y^{\prime})^{2})+a_{z}(z^{\prime})^{2}  
\label{S2}
\end{eqnarray}
with the angle of rotation fixed by the relation 
\begin{equation}
\tan 2\phi =-{\frac{a}{a_{x}-a_{y}}}\; .
\label{phi} 
\end{equation}
The deformation parameter (\ref{delta}), calculated in the rotating frame,
then takes the form (we assume $\delta \ge 0$) 
\begin{equation}
\delta ={\frac{\sqrt{(a_{x}-a_{y})^{2}+a^{2}}}{(a_{x}+a_{y})}} \, .
\label{delta2}
\end{equation}

The equations of motion for the nine parameters $a_{i}$, $\alpha _{i}$, 
$a$, $\alpha $, and $b$ are obtained by inserting Eqs.\ (\ref{rho_ansatz}
-\ref{v_ansatz}) into the hydrodynamic Eqs.\ (\ref{continuity}) and 
(\ref{force}).  The equation of continuity yields 
\begin{eqnarray}
\dot{b}+\left( \sum_{i=xyz}\alpha _{i}\right) b &=&0  \nonumber \\
\dot{a}_{x}+\left( \sum_{i=xyz}\alpha _{i}\right) a_{x}+2\alpha
_{x}a_{x}+\alpha a &=&0  \nonumber \\
\dot{a}_{y}+\left( \sum_{i=xyz}\alpha _{i}\right) a_{y}+2\alpha
_{y}a_{y}+\alpha a &=&0  \nonumber \\
\dot{a}_{z}+\left( \sum_{i=xyz}\alpha _{i}\right) a_{z}+2\alpha _{z}a_{z}
&=&0  \nonumber \\
\dot{a}+\left( \sum_{i=xyz}\alpha _{i}\right) a+\left( \alpha _{x}+\alpha
_{y}\right) a+2\alpha \left( a_{x}+a_{y}\right)  &=& 0\nonumber\\
\label{a_eqs}
\end{eqnarray}
while the equation for the force provides the additional five equations 
\begin{eqnarray}
\dot{\alpha}_{x}+\alpha _{x}^{2}+\alpha ^{2}-\left( \frac{2g}{m}\right)
a_{x} &=&0  \nonumber \\
\dot{\alpha}_{y}+\alpha _{y}^{2}+\alpha ^{2}-\left( \frac{2g}{m}\right)
a_{y} &=&0  \nonumber \\
2\dot{\alpha}+2\left( \alpha _{x}+\alpha _{y}\right) \alpha -\left( \frac{2g%
}{m}\right) a &=&0  \nonumber \\
\dot{\alpha}_{z}+\alpha _{z}^{2}-\left( \frac{2g}{m}\right) a_{z} &=& 0
\label{alpha_eqs}
\end{eqnarray}
Equations (\ref{a_eqs}) and (\ref{alpha_eqs}) complete the nine first--order
ordinary differential equations that we must solve in order to get $n$ and $%
{\bf v}$ as a function of time. 
An equivalent formalism was developed
in \cite{olshani} to investigate the behaviour of   nonlinear oscillations 
carrying  angular momentum.

Equations (\ref{a_eqs}) and (\ref{alpha_eqs}) ensure the conservation of
the integrals of motion, i.e., the number of atoms, $N$, the total energy 
$E$ and the angular momentum, $L_z$. These quantities can be expressed
in terms of our variables. For example, the integral of the density over 
all space must equal $N$: 
\begin{equation}
N=\frac{8\pi }{15}\frac{b^{5/2}}{\left( a_{x}a_{y}-a^2/4\right)
^{1/2}}a_{z}^{-1/2}   
\label{Na}
\end{equation}
while  the angular momentum, given by the irrotational law 
$L_z =Nm\delta^2\Theta_{rig}$, takes the form
\begin{equation}
L_z =\frac{1}{7}mN\Omega\delta^2
\frac{b(a_x+a_y)}{a_xa_y-a^2/4} \, .
\label{Lz}
\end{equation}
The conservation laws (\ref{Na}-\ref{Lz}) have been used to check the 
consistency of our numerical solutions.

\begin{figure}[!h]
\begin{center}
\mbox{\psboxto(3.4in;0in){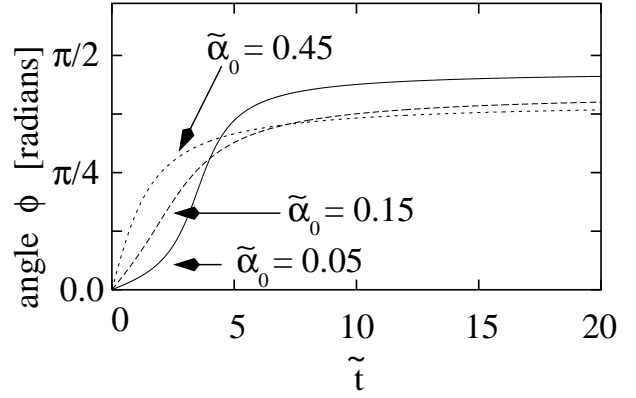}}
\end{center}
\caption{Angle of rotation of the condensate during the expansion as 
a function of time for different values of the initial angular velocity. 
The choice of the  parameters is given in the text. Time is given in 
dimensionless units (see text).}
\label{fig1}
\end{figure}

It is important to note that the equations of motion (Eqs.\ (\ref{a_eqs}) 
and (\ref{alpha_eqs})) are invariant with respect to the scaling
transformation 
$t\rightarrow t/\Lambda$, $\alpha_{i}\rightarrow\Lambda\alpha_{i}$, 
$\alpha\rightarrow\Lambda\alpha$, 
$a_{i}\rightarrow\Lambda^{2}a_{i}, a\rightarrow\Lambda^{2}a$, 
so that the initial size of the condensate can be actually excluded from 
the solution. For example, the dimensionless quantity $\delta $ can be 
presented as $\delta\left(\widetilde{t}\right)$, where 
$\widetilde{t}=t(2ga_{x0}/m)^{1/2}$. For small initial angular
velocities, $\widetilde{t}$ coincides with $\omega _{x}t$ where 
$\omega_x$ is the frequency of the trapping potential.  The function 
$\delta\left(\widetilde{t}\right) $ depends only on the initial shape 
of the condensate, i.e., on the ratios $a_{y0}/a_{x0}$, $a_{z0}/a_{x0}$ 
and on the dimensionless constant $\widetilde{\alpha }_{0}=\alpha_{0}
(2ga_{x0}/m)^{-1/2}$. The same behavior holds for the angle of rotation 
$\phi\left( \widetilde{t}\right) $ and for the dimensionless angular 
velocity $\widetilde{\Omega}\left(\widetilde{t}\right) =
d\phi/d\widetilde{t}=\Omega(2ga_{x0}/m)^{-1/2}$.

\begin{figure}[!h]
\begin{center}
\mbox{\psboxto(3.4in;0in){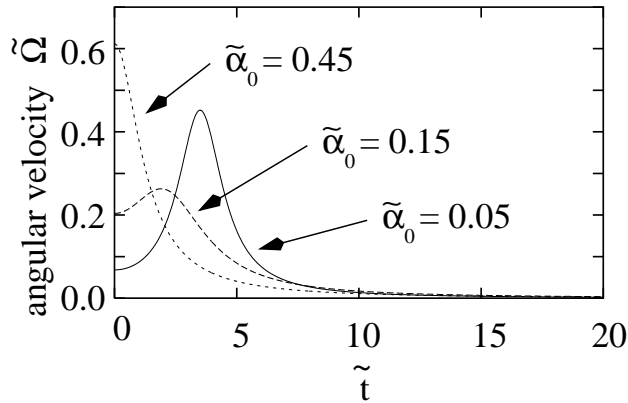}}
\end{center}
\caption{Angular velocity of the condensate during the expansion as a 
function of time for different values of the initial angular velocity 
(see Fig.\ 1.)}
\label{fig2}
\end{figure}

\begin{figure}[!h]
\begin{center}
\mbox{\psboxto(3.4in;0in){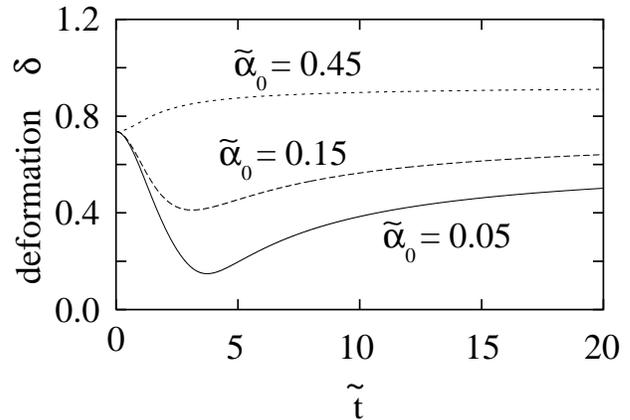}}
\end{center}
\caption{Deformation parameter of an expanding condensate
as a function of time for different values of the initial angular
velocity (see Fig.\ 1).}
\label{fig3}
\end{figure}

The results of the numerical integration of Eqs.\ (\ref{a_eqs}) and 
(\ref{alpha_eqs}) are presented in Figs.\ \ref{fig1}-\ref{fig3}. We have 
assumed that initially the condensate has a cigarlike shape elongated 
along the $y$--direction with $a_{z0}/a_{x0}=1,a_{y0}/a_{x0}=\lambda^{2}$ 
and $\lambda = 0.39$.  This corresponds to an initial deformation in the 
$x-y$ plane given by $\delta_{0}=(1-\lambda^{2})/(1+\lambda^{2})=0.74$. 
Time is measured in units of the dimensionless quantity $\widetilde{t}$. 
Three different values of the parameter $\widetilde{\alpha }_{0}$ have 
been considered ($0.05, 0.15, 0.45$), corresponding to different choices 
of the initial angular velocity and of the trapping parameters. 
In the presence of rotation the relationships between the deformation 
parameters of the trap and the ones  of the condensate are not trivial.

Using the stationary solutions derived in \cite{RZS} we find that the above
choices for $\tilde{\alpha}_0$ and $\lambda$ correspond to the values 
$\Omega_0 = 0.07\omega_x,\ \omega_y = 0.39\omega_x,\ \omega_z=1.0\omega_x$;
$\Omega_0 = 0.21\omega_x,\ \omega_y= 0.45\omega_x,\ \omega_z=1.0\omega_x$
and $\Omega_0 = 1.2\omega_x,\ \omega_y = 1.41\omega_x,\ \omega_z=2.0\omega_x$
respectively.  Notice that, except for values of $\Omega_0$ much smaller 
than the trapping frequencies $\omega_x, \omega_y$, the relationship 
between $\alpha_0$ and $\Omega_0$ is not linear since the parameters 
$a_{x0}, a_{y0}$ entering Eq.\ (\ref{alpha0}) are renormalized by the 
rotation of the trap. In particular  the last choice ($\Omega_0 = 1.2
\omega_x,\ \omega_y = 1.41\omega_x$) corresponds to a configuration 
belonging to the overcritical branch discussed in~\cite{RZS} where the 
deformation of the condensate has opposite sign with respect to the one 
of the confining trap. 

Figure \ref{fig1} shows that the angle of rotation $\phi\left(\widetilde{t}
\right)$ reaches, for large times, a constant value.  This is due to 
the rapid increase in the condensate size during the expansion which,
in turn, causes a fast decrease of the angular velocity.  The actual 
value of the asymptotic angle depends on the initial conditions, 
differently from what happens in the ideal gas where $\phi$ always 
approaches the value $\pi/2$.   In Fig.\ \ref{fig2} we show the 
angular velocity $\widetilde{\Omega}\left( \widetilde{t}\right)$.  
One can see that, for small values of the initial angular velocity, 
$\widetilde{\Omega }$ exhibits a clear enhancement in the first stage 
of the expansion.  Correspondingly the deformation parameter $\delta
\left( \widetilde{t}\right)$ reaches a minimum (see Fig.\ \ref{fig3}).  
If the initial angular velocity is large, the minimum is shallow and 
eventually disappears. In this case also the angular velocity changes 
its behavior and decreases monotonically with time (see dotted line in 
Fig.\ \ref{fig2}).  In the recent experiment of Ref.\ \cite{ens01}, steady 
rotating states of magnetically trapped condensates were generated using 
rotating laser beams. The condensates were then imaged after expansion.  
By using the formalism of ref.\cite{olshani} the authors of 
Ref.\ \cite{ens01} have checked that the deformation of the expanding 
condensate exhibits only minor changes. This result is the consequence of 
the relatively large values of the initial angular velocity considered 
in this experiment.

In conclusion we have shown that the expansion of a rotating condensate 
reveals dramatic consequences of superfluidity associated with the 
reduction of the moment of inertia. These effects appear as a sudden 
increase of the angular velocity accompanied by the occurrence of a 
minimum  in the deformation parameter. These effects, which are 
particularly pronounced for small initial angular velocities, should be 
easily observable by taking consecutive images during the expansion
of the rotating condensate. 

Useful discussions with Francesca Zambelli and David Feder are 
acknowledged.  M.E.\ and C.W.C.\ acknowledge partial support from NSF
grants nos.\ 9802547 and 9803377.  This work was supported by the 
Ministero della Ricerca Scientifica e Tecnologica (MURST).


\end{document}